\documentclass[9pt,twocolumn,twoside]{opticajnl}
\journal{opticajournal} %

\setboolean{shortarticle}{false}

\usepackage{lineno}
\usepackage{sidecap, caption}
\usepackage{sidecap}
\usepackage{wrapfig}
\usepackage{graphicx}
\usepackage{bm}
\usepackage{ulem}  %% REMOVE ONCE \sout{} is no longer used
\usepackage{braket}
\usepackage{comment}
\DeclareMathOperator{\sinc}{sinc}
\title{Ultrashort-pulse-pumped, single-mode type-0 squeezers in lithium niobate nanophotonics}

\vspace{-2mm}

\author[1]{Martin Houde}
\author[2]{Liam Beaudoin}
\author[2]{Robert Kwolek}
\author[2,5]{Kazuki Hirota}
\author[2,3,4]{Rajveer Nehra}
\author[1]{Nicol\'as Quesada}
\vspace{-3mm}

\affil[1]{Department of Engineering Physics, \'{E}cole polytechnique de Montr\'{e}al, Montr\'{e}al, Qu\'{e}bec, H3T 1J4, Canada}
\affil[2]{Department of Electrical and Computer Engineering, University of Massachusetts Amherst, Amherst,
Massachusetts 01003, USA}
\affil[3]{Department of Physics, University of Massachusetts Amherst, Amherst,
Massachusetts 01003, USA}
\affil[4]{College of Information and Computer Science, University of Massachusetts Amherst,
Amherst, Massachusetts 01003, USA}
\affil[5]{Department of Applied Physics, The University of Tokyo, 7-3-1 Hongo, Bunkyo-ku, Tokyo 113-8656, Japan}

\affil[$^\dagger$]{rajveernehra@umass.edu, $^*$nicolas.quesada@polymtl.ca}

\begin{abstract}
We present design principles for ultrashort-pulse, type-0 phase-matched optical parametric amplifiers to generate and measure spectrally pure degenerate squeezed light. {We consider a fundamental signal (second-harmonic) mode at 2090 (1045) nm} and show that our proposed design achieves a Schmidt number of $K \approx 1.02$ with squeezing levels greater than 15 dB on a single temporal mode spanning over $5$ THz in bandwidth with cm-scale devices on thin-film lithium niobate (TFLN) on insulator platform. Our work opens up promising avenues for large-scale circuits for ultrafast quantum information processing and quantum sensing applications on the rapidly advancing TFLN platform with already demonstrated linear components and photodetection capabilities. 
\end{abstract}

\setboolean{displaycopyright}{false} %

\begin{document}

\maketitle
\section{Introduction}

Integrated photonics is a platform for scalable quantum information systems (QISs)~\cite{pelucchi2022potential,Arrazola2021,bourassa2021blueprint,aghaee2025scaling,maring2024versatile,psiquantum2025manufacturable}. A squeezed light source that meets the following criteria is a fundamental building block for photonic quantum technologies~\cite{vernon2019scalable}: \textit{(i)} Integration of multiple identical, coherent, and stabilized sources for complex quantum circuits, \textit{(ii)} Squeezed light in a consistent single spatiotemporal mode across various squeezing levels, eliminating narrow-band filters, \textit{(iii)} Squeezing levels enabling quantum advantage in information processing, sensing, and metrology, and \textit{(iv)} Compatibility with ultra-broadband fields and photon counting measurements. Besides this one could also add \textit{(v)} the ability to perform perform inline squeezing, thus using the source as a broadband amplifier for measurement

Integrated squeezers using spontaneous four-wave mixing (mediated by a third-order nonlinear susceptibility $\chi^{(3)}$) in SiN and silica resonators~\cite{vernon2019scalable} meet desiderata \textit{(i-iv)} but need precise filtering and locking due to proximity of pump-signal resonance and frequency sensitivity to thermal fluctuations. Alternatively, single-pass squeezers based on phase-sensitive optical parametric amplifiers (OPAs) utilizing stronger second-order optical nonlinearity $\chi^{(2)}$ have been proposed and demonstrated in continuous-wave (CW) and pulsed regimes~\cite{kashiwazaki2023over, Nehra2022few,quesada2018gaussianA,roman2024multimode}. 
In such OPAs, the spontaneous parametric downconversion (SPDC) process allows the signal photons to be emitted across multiple modes governed by the pump profile and phase-matching conditions. Ultrashort-pulsed pump OPAs, in particular, produce photons across numerous spatiotemporal modes, which reduces the quantum state purity for heralded single-photon and squeezed light generation applications. While spectrally de-correlated single-mode operation is achievable in non-degenerate type-II twin-beam squeezing through careful design~\cite{uren2005pure,mosley2008heralded, quesada2018gaussianA, houde2022waveguided, houde2024perfect}, the design principles for single-mode degenerate squeezing with type-0 SPDC sources remain undeveloped. Although twin-beam squeezing can, in principle, be converted to degenerate squeezed modes using linear optics (e.g., beam splitters and mode converters), this approach can introduce additional losses and complexities, failing to meet \textit{i--iii}.

In this work, we introduce design principles for ultrashort-pulse-pumped, near-single spatiotemporal mode travelling-wave squeezers using type-0 phase matching, engineered through the group velocity and group velocity dispersion of the relevant modes on a rapidly emerging thin-film lithium niobate platform~\cite{zhu2021integrated}. {For a fundamental signal at 2090 nm,  in the ``single photon-pair generation (SPG)} limit'', our design can achieve a Schmidt number of $K \approx 1.115$ for a $50$ fs pump and generate squeezing with a main broadband mode spanning $5$ THz in bandwidth. This further improves to about $K\approx 1.02$ in the high-gain regime. %
Additionally, we demonstrate the efficacy of these OPAs in the high-gain regime for all-optical, loss-tolerant, ultra-broadband quadrature measurements %
	, achieving near-unity fidelity between the high-gain measurement OPA (MOPA) mode and the system (squeezer OPA + MOPA) (as discussed in Sec.~\ref{sec:all_optical}).
	Our designs satisfy desiderata \textit{(i–v)}, thereby paving a practical path for ultrafast information processing and sensing applications. 

This manuscript is structured as follows: in Sec.~\ref{sec:principles} we introduce our design principles in the single-photon-pair generation regime and show that they lead to excellent operation also in the high-squeezing regime. In Sect.~\ref{sec:tfln} we show how our design principles can be implemented in a thin-film lithium niobate on an insulator platform in the mid-infrared and show that our proposed implementation is resilient to manufacturing imperfections. In Sec.~\ref{sec:sup} we show that we can suppress processes in other spatial modes for the proposed design and thus guarantee control in all the relevant degrees of freedom in our proposed design. In Sec.~\ref{sec:all_optical} we show how our design can also be used in a so called all-optical loss-tolerant measurement device. In Sect.~\ref{sec:telecom} we show that our design principles can also be implemented in the C- and L-bands of fiber optic communication, albeit with more restrictive manufacturing specifications. Finally, the conclusion and outlook follow in Sect.~\ref{sec:end}.

\section{Design principles}\label{sec:principles}

A general squeezed state can be written as 
\begin{align}
\ket{\text{sq}[J]}=\exp\left(\tfrac12 \int d\omega  d\omega' \left[ J(\omega, \omega') a^\dagger(\omega) a^\dagger (\omega') - \text{H.c.}\right]\right) \ket{\text{0}}.
\end{align} Here, $\ket{\text{0}}$ is the vacuum state, $a^\dagger(\omega)$/$a(\omega)$ are bosonic creation/annihilation operators at frequency $\omega$ with canonical commutation relation $[a(\omega),a^\dagger(\omega')] = \delta(\omega-\omega')$, $J(\omega, \omega') = J(\omega', \omega)$ is the symmetric joint spectral amplitude (JSA) defining $\ket{\text{sq}[J]}$, "H.c." stands for the Hermitian conjugate, and integration spans positive frequencies. Spectral separability requires $\left. J(\omega,\omega')\right|_{\text{separable}} =  r f(\omega) f(\omega')$ where $f(\omega)$ is $L^2$ normalized and $r$ is the squeezing parameter, with corresponding decibel level squeezing given by $10 \log_{10} e^{2r} $. Requirements \textit{(i-ii)}  mathematically translate to a separable JSA at varying squeezing levels for a given frequency mode $f(\omega)$. The Schmidt number characterizes the single-modeness of the state, defined as
\begin{align}
	K[J] =  \left[\sum_{k} \sinh^2 r_k \right]^2 / \sum_k \sinh^4 r_k.
\end{align}
If two {out-of-phase} copies of the degenerate squeezed state can be mixed on a 50:50 symmetric beam splitter, {resulting in a two-mode squeezed state,} then the quantity above can be inferred directly from an unheralded second-order correlation measurement~\cite{christ2011probing}. 
The squeezing parameters $r_k$ are obtained by writing the Takagi~\cite{arzani2018versatile,houde2024matrix,horoshko2024few,quesada2022beyond} decomposition $J(\omega,\omega') = \sum_{k} r_k f_k(\omega) f_k(\omega') $), with $K = 1$, implying $r_k = r\delta_{k,0}$ and indicating a single-mode squeezer.

In writing the expression for $\ket{\text{sq}[J]}$ we are assuming that the transverse degrees of freedom are controlled by the waveguide and that no other waveguided modes are phase-matched (which we confirm numerically for the lithium niobate on insulator in Sec.~\ref{sec:all_optical} ).
Hence, as we will show, our design achieves a single spatiotemporal mode by making sure that no other spatial mode can undergo nonlinear wave mixing (thus controlling the transverse degrees of freedom of the process) and also controlling the frequencies generated (thus also controlling the longitudinal parts).

\begin{figure}[!t]%
	\centering
	\includegraphics[width = 1\linewidth]{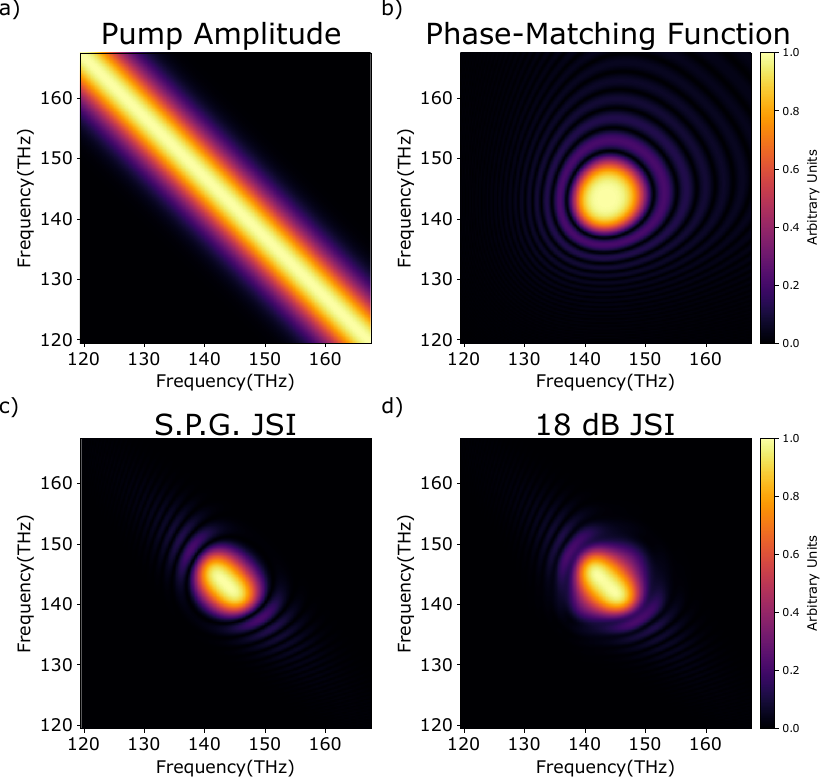}
	\caption{(a) Pump spectral amplitude for a 50 fs pulse. (b) Phase-matching function. (c) JSI in the SPG regime. (d) JSI in the high-gain limit for 18 dB of gain.}
	\label{fig:jsi}
\end{figure}

In type-0 configuration with second harmonic (SH) pump (P, $2 \bar \omega$) and fundamental harmonic (FH) signal (S, $ \bar \omega$) modes, the JSA for nearly single-mode {operation and for} $<10$ dB squeezing is~\cite{helt2020degenerate}
\begin{align}\label{eq:jsa}
J(\omega,\omega') = C \alpha(\omega+\omega'-2 \bar{\omega}) \sinc \tfrac{ \ell}{2} \Delta k(\omega, \omega'), 
\end{align} where $\alpha(\omega + \omega'-2 \bar{\omega})$ is the pump amplitude evaluated at the sum of the frequencies of the pair of fundamental photons (relative to the central frequency of the pump $2\bar{\omega}$), $\ell$ is the OPA length, {$\sinc \tfrac{ \ell}{2} \Delta k(\omega, \omega')$ is the phase-matching function (PMF)}, and $C$ is a normalization constant. The phase mismatch $\Delta k(\omega, \omega') = - k_{{P}}(\omega'+\omega)+ k_S(\omega)+k_S(\omega')$ can then be simplified using dispersion relations as,
\begin{align}\label{eq:pm}
	\Delta k(\omega, \omega') =&
	- \bar{k}_P - \tfrac{1}{v_P}(\delta \omega + \delta \omega') - \tfrac12 \beta_P  (\delta \omega+\delta \omega')^2 \\
	&+  2 \bar{k}_S + \tfrac{1}{v_S}(\delta \omega + \delta \omega') + \tfrac12 \beta_S [(\delta \omega)^2+(\delta \omega')^2]  , \nonumber
\end{align}
where $\delta \omega = \omega - \bar{\omega}$ and $\beta_{S}/\beta_{P}$ are the group velocity dispersion (GVD) terms. We will assume that the terms independent of frequency in the equation above, namely $2 \bar{k}_S - \bar{k}_P$, add up to 0  or  to $\pm 2\pi /\Lambda$ if poling with period $\Lambda$ is used for quasi-phasematching. When the group velocity matching (GVM) condition $v_P \approx v_S$ is satisfied, and moreover $\beta_S \gg \beta_P$ we can approximate the JSA as
\begin{align}
	J(\omega,\omega') \approx C \alpha(\omega+\omega'-2 \bar{\omega}) \ \sinc \tfrac{\ell \beta_S}{4} [(\delta \omega)^2+(\delta \omega')^2],  
\end{align}
Furthermore, if the $\sinc$ function, which roughly occupies a circular area of radius $2/\sqrt{\beta_S L}$ in the $(\omega,\omega')$ space is well within the bandwidth of the pump (corresponding to a band at -$45^\circ$ in $(\omega,\omega')$ space) the we can approximate 
\begin{align}
	J(\omega,\omega') \approx C' \sinc \tfrac{\ell \beta_S}{4} [(\delta \omega)^2+(\delta \omega')^2], \  C'=C \alpha(0) ,
\end{align}
which, {to a good approximation}, is a separable function in its two arguments. {We have minimized the overlap between the pump amplitude and the PMF by effectively cancelling out the leading order terms in the phase mismatch that are proportional to powers of $(\delta \omega + \delta \omega')$. These terms are parallel to the pump band in $(\omega,\omega')$ space and, when present, lead to high overlap with the pump and low spectral purity. By demanding that $\beta_S \gg \beta_P$, we also ensure that the dominating term in the phase mismatch is rotationally symmetric and concentrated near central frequency $\omega = \omega' = \bar{\omega}$. This gives rise to a PMF that is also rotationally symmetric and decays quickly past its characteristic bandwidth $\tfrac{2}{\sqrt{\beta_S \ell}}$. In Fig.~\ref{fig:jsi}(a), we show a typical pump spectral amplitude for our interactions. Fig.~\ref{fig:jsi}(b) shows the PMF for the case where these conditions are met. Combining both gives us the joint spectral intensity (JSI) shown in Fig.~\ref{fig:jsi}(c).}

While the simple analytical theory supporting the design described above is only valid in the single-photon pair limit regime {and up to roughly 10 dB of squeezing}, we perform quantum-nonlinear optical simulations~\cite{helt2020degenerate,quesada2022beyond} of the problem valid for arbitrary levels of gain (modulo pump depletion, which is a valid approximation in all that we present below). {For these high-gain simulations, we set the pump {spectral} profile to be Gaussian for a given length of the waveguide, and vary the level of gain by tuning the pump power, while keeping the pump \emph{spectral profile fixed}. These simulations take into account time-ordering corrections and output the second-order moments $N_{S}(\omega,\omega') =\langle \text{sq}[J]| a^{\dagger}(\omega)a(\omega') |\text{sq}[J] \rangle $ and $M(\omega,\omega') =\langle \text{sq}[J]| a(\omega)a(\omega') |\text{sq}[J] \rangle$ where $\ket{\text{sq}[J]}$ is the squeezed state for arbitrary gain. We construct the high-gain JSA from the singular-value decomposition of the $M(\omega,\omega')$ moment~\cite{Quesada2020-pdc}. This also gives us the resulting squeeze parameters $r_{k}$ from which we can obtain the Schmidt number, allowing us to verify the near-single modeness of the source. Moreover, we show that it does not degrade (in fact, it improves) as the gain increases, as shown in Fig.~\ref{fig:schmidt_combined}(a). From the eigendecomposition of $N(\omega,\omega')$, we can obtain the temporal mode structure of the longitudinal modes, allowing us to find the operational bandwidth. We determine the overall level of gain from the average number of signal photons.}
\begin{align}\label{eq:averageN}
    \langle N_{S}\rangle = \int d\omega \ N_{S}(\omega,\omega) = \sum_{k =0}^{\infty} \sinh^2 r_k.
\end{align}

\begin{figure}[!t]%
	\centering
	\includegraphics[width = 1\linewidth]{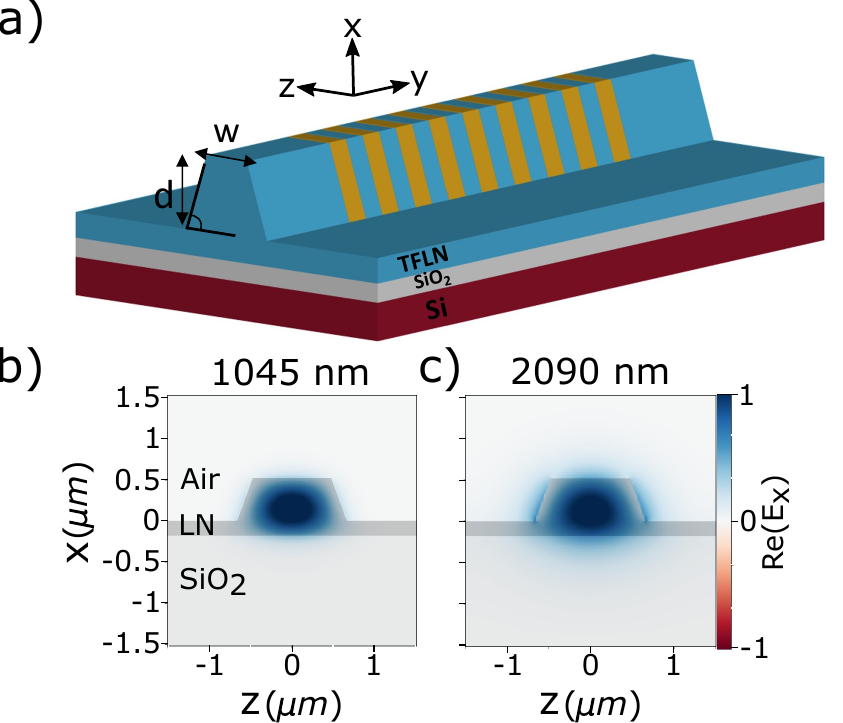}
	\caption{(a) Periodically-poled, x-cut thin-film ($700~\mathrm{nm}$) lithium niobate on insulator ridge waveguide with top width $w$ and etch depth $d$, supported on a $2~\mu\mathrm{m}$ $SiO_2$ layer and a $500~\mu\mathrm{m}$ Si substrate. (b) Transverse mode profiles for SH and FH modes, showing the tight mode confinement.}
	\label{fig:geometry}
\end{figure}

{The average photon number $\langle N_{S}\rangle$ depends implicitly on the pump power. By tuning the pump power, we can realize arbitrary values of $\langle N_{S}\rangle \equiv \sinh^{2}(r_{\text{eff}})$ for an effective squeezing parameter $r_{\text{eff}}$, which can then be expressed in decibels. In Fig.~\ref{fig:jsi}(d) we show the JSA for an effective squeezing level of 18 dB corresponding to a mean photon number of $\langle N_{S}\rangle \approx 15.3$.
To further quantify single-modeness, we define the mode occupation fraction for the $k$-th mode,}
\begin{align}\label{eq:mode_frac}
	\text{Mode fraction} (k) =\frac{\sinh^2 r_k}{\langle N_{S}\rangle}, 
\end{align}
{where $r_k$ is the squeezing parameter for $k$-th mode. }

\section{Lithium niobate nanophotonics}\label{sec:tfln}

We design OPAs on the emerging thin-film lithium niobate on insulator platform~\cite{ledezma2022intense,Nehra2022few}, shown in Fig.~\ref{fig:geometry}(a). In recent years, thin-film lithium niobate (TFLN) on insulator has emerged as a promising platform to address these outstanding challenges in quantum nanophotonics~\cite{boes2023lithium, vazimali2022applications,wang2018integrated}. In TFLN, the synergy of sub-wavelength mode confinement, strong second-order ($\chi^{(2)}$) optical nonlinearity, efficient quasi-phase-matching (QPM) through periodic poling, and dispersion engineering for prolonged spatiotemporal confinement has led to devices outperforming traditional bulk LN devices by more than an order of magnitude in nonlinear efficiency and bandwidth. The stronger $\chi^{(2)}$ allows one for single-pass travelling-wave nonlinear quantum device waveguides, circumventing technical challenges in high-Q resonators\cite{Nehra2022few,  Rao2019, ledezma2022intense, guo2022femtojoule,Jankowski2020}. Moreover, larger electro-optic coefficients in TFLN have enabled electro-optic modulators (EOMs) with a bandwidth exceeding 100 GHz, reduced $V_{\pi}$ voltage, and a small device footprint for large-scale programmable interferometers {to route quantum fields on a chip for quantum computing, communication, and sensing applications~\cite{wang2018integrated,Juneghani2023}. In particular, the ultra-high bandwidth supports fast clock rates for photonic quantum processors. Collectively, these features position TFLN as a promising platform for scalable quantum photonics.}

Considering {FH} and SH modes at $2090$ nm and $1045$ nm (mode profiles in Fig.~\ref{fig:geometry}(b),(c)) with a 700 nm thin-film thickness and $\theta = 70^{\circ}$ ridge angle, we find optimal parameters: etching depth $d = 517.7$ nm and top width $w = 942.3$ nm through numerical simulations~\cite{snow2024}. These satisfy GVM ($v_P - v_S \simeq 0$) and GVD ($\beta_S = 650~\text{fs}^2/\text{mm} \gg \beta_P = 50~\text{fs}^2/\text{mm}$) conditions for a 50 fs SH pump and 10-mm-long device.

\begin{figure*}
	\centering
	\includegraphics[width = 1\linewidth]{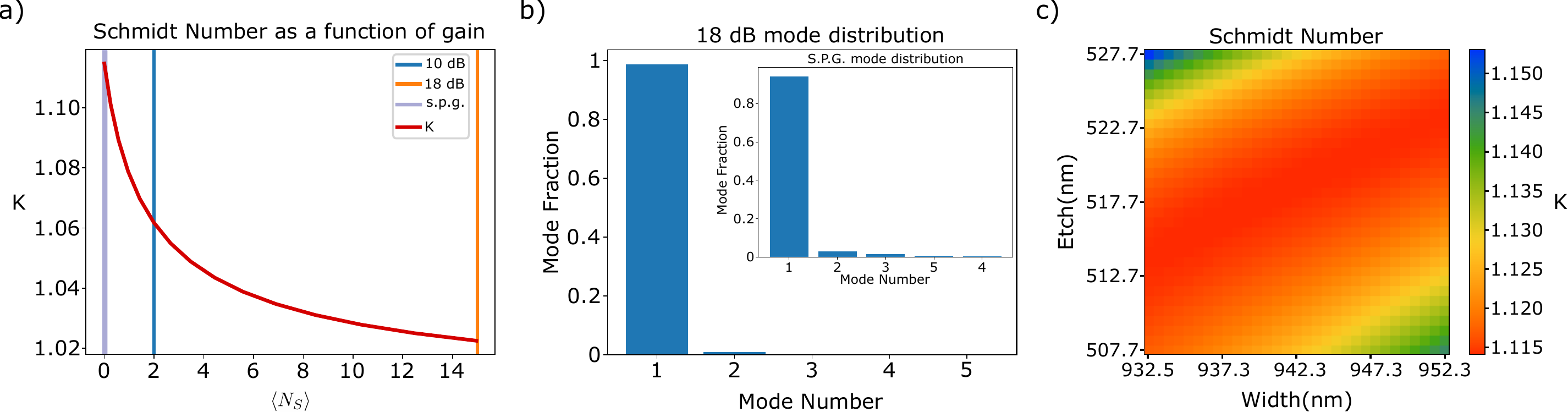}
	\caption{(a) Schmidt number as a function of gain, determined by the average number of signal photons ($\langle N_{S}\rangle$, c.f. Eq.~\ref{eq:averageN}). Vertical lines represent different levels of gain in dB. The Schmidt number varies from $K\approx 1.115$ (SPG regime) to $K\approx 1.02$ (18 dB gain limit). (b) Mode occupation fraction (c.f. Eq.~\ref{eq:mode_frac}) for the first five modes in the 18 dB gain limit (inset: SPG regime). We find a first/second mode occupation ratio of 0.989/0.007 (0.95/0.02) in the 18 dB gain limit (SPG regime). (c) Schmidt number in the SPG regime as a function of deviations from the ideal geometry. The waveguide width and etch depth are varied by $\pm 10~\mathrm{nm}$ while keeping the film thickness and ridge angle fixed. The Schmidt number increases slightly away from the ideal geometry but remains close to unity, showing the fabrication tolerance of our proposed designs.}
	\label{fig:schmidt_combined}
\end{figure*}

{In Fig.~\ref{fig:jsi}(c) and (d) we show the JSI for the SPG and high-gain limits respectively. Although both look fairly similar, the central area of the SPG JSI is slightly elliptical, whereas,  in the high-gain limit, time-ordering effects distort the central area, making it slightly more circular. This, combined with the exponentially faster growth of the first temporal mode, compared to the others (i.e. $r_{1}\gg r_{i\neq1}$), allows us to achieve near-separability (Schmidt number $K \sim 1$). }

{Indeed, in Fig.~\ref{fig:schmidt_combined}(a) we show how the Schmidt number varies as a function of the average signal photon number, spanning SPG to high squeezing. We see that the Schmidt number varies from $K\approx 1.115$ in the SPG limit to $K\approx 1.02$ at 18 dB gain. In Fig.~\ref{fig:schmidt_combined}(b) we show the mode occupation fractions, defined in Eq.~\ref{eq:mode_frac}, for the first five temporal modes at 18 dB (inset: SPG regime). We find a first/second mode ratio of 0.989/0.007 (0.95/0.02), demonstrating the versatility of our design for single-photon and squeezed light generation.}

To show the robustness of our design to fabrication uncertainties/errors, we consider how the purity degrades in the SPG regime when we allow for variations in the width and edge depth parameters. In Fig.~\ref{fig:schmidt_combined}(c), we vary both parameters by $\pm 10$ nm, which is on par with fabrication limitations~\cite{xin2025wavelength, boes2023lithium,zhu2021integrated,renaud2023sub}. {Although the Schmidt number increases slightly, it does so on the order of 0.04 which is small and does not greatly affect the single-modeness.}

%We see that the Schmidt number increases slightly. However, these increases are small, on the order of 0.04, and do not greatly affect the single-modeness.

\section{Suppression of parasitic processes by phase-mismatching}\label{sec:sup}

{Careful control of spatial modes is crucial to suppress unwanted parametric interactions. The concurrent generation of two-mode squeezing (TMS) between the fundamental transverse electric (TE) and transverse magnetic (TM) modes, along with the desired single-mode squeezing in the TE mode, can degrade the observed squeezing when measurements are restricted to the fundamental TE mode. Although such effects can be alleviated by optimizing the measurement basis through spatial shaping of the local oscillator--it incurs substantial hardware overhead, bandwidth limitations imposed by spatial light modulators, and increased experimental complexity~\cite{sano2023effects}. }\\
{Our approach addresses this by ensuring that higher-order modes and cross-polarized fundamental TM modes are strongly phase mismatched for cm-scale OPAs considered here, that is, $\Delta \bar{k}$ is sufficiently large for the chosen poling period so that the effective phase mismatch $\Delta \bar{k} \ell$ exceeds the threshold set by the parametric gain. We calculate the phase mismatch $\Delta \bar{k}=\Delta \bar{k}'- \frac{2\pi}{\Lambda_{\text{QPM}}}$ for a number of waveguide geometries to account for fabrication imperfections for the geometry considered. Here 
	\begin{align}
		\Lambda_{\text{QPM}} = \frac{2\pi}{k_{\text{SH,TE}_0} - 2k_{\text{FH,TE}_0}},
	\end{align}
	is the poling period such that $\Delta \bar{k}=0$ corresponding to a quasi-phase-matched degenerate interaction for the fundamental TE modes of a given geometry. We perform numerical simulations by sweeping the top width from $922~\mathrm{nm}$ to $962~\mathrm{nm}$ and the etch depth from $497.8~\mathrm{nm}$ to $537.8~\mathrm{nm}$, both in $10~\mathrm{nm}$ increments. The sidewall angle is kept fixed at $70^\circ$. For each waveguide geometry, we compute the effective refractive indices for the first 20 eigenmodes using Tidy3D’s mode solver~\cite{tidy3d}. To identify guided modes, we calculate the \emph{confinement ratio}, defined as the ratio of the time-averaged Poynting vector $\langle S_y\rangle$ integrated over the ridge to the total power in the simulation domain. Modes with a confinement above 70\% are classified as guided. We then calculate the phase mismatch $ \Delta \bar{k} $ for all valid combinations of guided modes at each geometry, excluding the fundamental TE interaction corresponding to $\Delta \bar{k} = 0$.}

\begin{figure*}[h]
	\centering
	\includegraphics[width=1\linewidth]{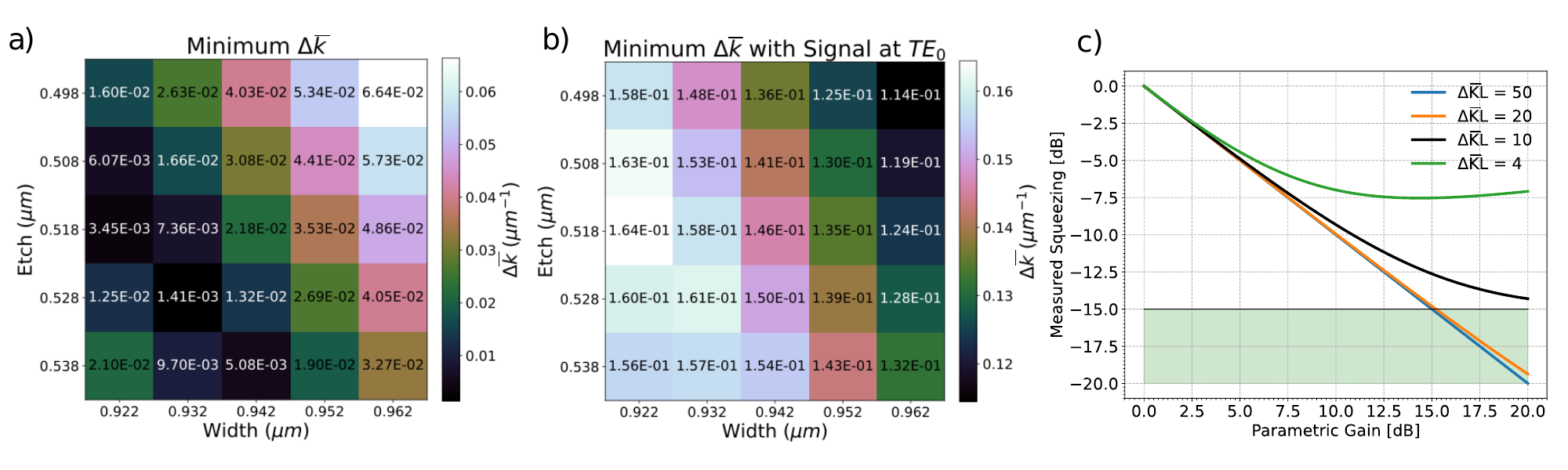}
	\caption{(a) Minimum phase mismatch, $\Delta \bar{k}$ from the unwanted interactions between higher-order spatial modes, evaluated across a range of waveguide geometries and modes above the set threshold for guided modes. The top width and etch depth are varied in 10 nm steps within $\pm20$ nm of the ideal design, while the film thickness and sidewall angle are held constant. (b) Minimum phase mismatch, $\Delta \bar{k}$ when the signal mode (i.e., desired mode for squeezing) is fixed as the fundamental TE mode while pump and idler modes are swept.
		(c) Impact of phase mismatch on squeezing for the ideal geometry, shown as a function of parametric gain (set by pump power) for different values of effective mismatch $\Delta \bar{k}L$. All geometries within the $\pm 20$ nm from ideal achieve $\Delta \bar{k} \ell > 20$, ensuring robustness to fabrication errors.}
	\label{fig:squeeze_deg}
\end{figure*}

{Fig.~\ref{fig:squeeze_deg}(a) presents the numerically simulated phase mismatch parameter $\Delta \bar{k}$, where the solver evaluates the waveguide modes at both the FH and SH wavelengths and identifies the worst-case scenario corresponding to the minimum $\Delta \bar{k}$. For the ideal design, we get $\Delta \bar{k} \ell > 200$, ensuring optimal phase matching in the desired modes. 
Likewise, Fig.~\ref{fig:squeeze_deg}(b) shows the minimum $\Delta \bar{k}$ when the signal is restricted to the fundamental $TE_0$ mode, where squeezing is targeted. Suppose that squeezing occurs in two modes between the $TE_0$ signal mode and the higher-order idler modes. In that case, it can introduce excess thermal noise into the single-mode (degenerate) squeezed vacuum state--unless additional measures, such as spatial shaping of the local oscillator, are employed in the measurement chain.

Fig.~\ref{fig:squeeze_deg}(c) illustrates how the phase mismatch affects measurable squeezing as a function of parametric gain in the fundamental TE pump mode. We assume no other noise sources (e.g., propagation loss and phase noise) here. 
Measured squeezing without a shaped local oscillator significantly reduces for moderate phase mismatch, with a noticeable loss of measured squeezing at gains as low as 3 dB for small phase mismatch $\Delta \bar{k} \ell = 4$. To achieve squeezing levels exceeding 15 dB in centimeter-scale OPAs, one needs a phase mismatch of $\Delta \bar{k}\ell> 20$ (i.e., $\Delta \bar{k} > 0.002~\mu\text{m}^{-1}$) --- a condition satisfied by {the majority of our designs considered in $\pm$20 nm}. The ideal waveguide geometry with etch depth $d = 517.7$ nm and top width $w = 942.3$ nm exhibits a significantly larger $\Delta \bar{k}$, while maintaining a fabrication tolerance of $\pm$20 nm around the ideal waveguide width and etching depth values. {Importantly, when the fundamental is restricted to $TE_0$ mode, the phase mismatch $\Delta k$ is an order of magnitude higher than the required minimum, as per Fig.~\ref{fig:squeeze_deg}(b).} It should be noted that optimized designs that suppress unwanted parametric interactions will also enhance the performance of all-optical measurements using high-gain OPAs.}

\begin{figure}[!t]%
	\centering
	\includegraphics[width = 1\linewidth]{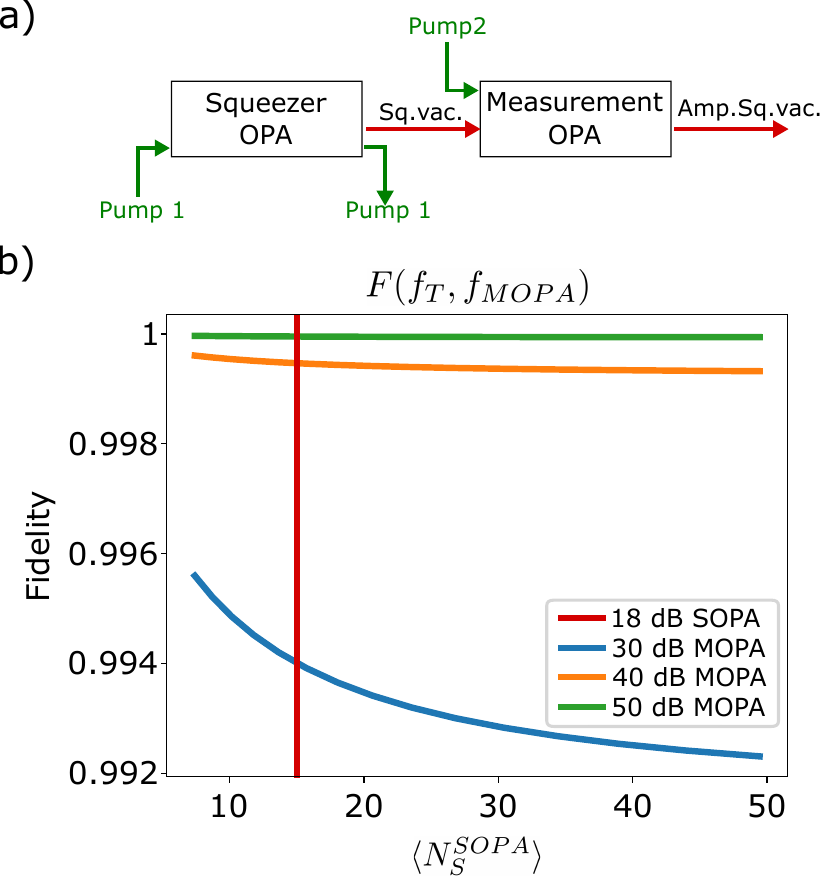}
	\caption{(a) Schematic of the all-optical, loss-tolerant measurement protocol. A squeezed vacuum (Sq. vac.) is generated in the SOPA using pump 1, and subsequently injected into the MOPA along with pump 2. The combined output is an amplified squeezed vacuum (Amp. Sq. vac.), which is then characterized through homodyne detection.
		(b) Fidelity [Eq.~\ref{eq:fidelity}] between the overall output temporal mode of the system ($f_{T}$) and the MOPA’s output temporal mode ($f_{MOPA}$), plotted as a function of the SOPA gain, quantified by the average number of signal photons $\langle N^{SOPA}_{S}\rangle$. Different curves correspond to varying MOPA gain levels. The red vertical line denotes the SOPA operating point at 18 dB gain. }
	\label{fig:mopa}
\end{figure}

\section{All-optical, loss-tolerant measurements %
	in the high-gain regime}\label{sec:all_optical}
While so far we have only discussed the utility of our design as a squeezed light source, we now consider its utility as a measurement optical parametric amplifier (MOPA) corresponding to desiderata \emph{(v)} in the introduction~\cite{Nehra2022few, takanashi2020all, shaked2018lifting, Kalash2022,kawasaki2024broadband}. 
We show the efficacy of our proposed OPA designs in the high-gain regime to perform recently emerging all-optical, loss-tolerant measurements for the $5$ THz bandwidth mode that we can generate. Such measurements will be a critical component for all-optical, ultra-fast quantum processors and networks~\cite{yamashima2025all}. 
The entire system (T, for total) consists of a squeezer OPA (SOPA) and a high-gain measurement OPA (MOPA), as shown in Fig.~\ref{fig:mopa}(a). In multi-mode systems, the modes can mix, leading to output Schmidt modes for the total system that may differ from the MOPA modes, thereby degrading the measurement quality. 

We can describe a system of two cascaded squeezers, such as a SOPA followed by a MOPA, using three different sets of input/output Schmidt modes: one set for each stage (SOPA modes and MOPA modes) and a third set representing the overall interaction. The MOPA serves as an anti-squeezer and typically operates with significantly higher gain than the SOPA. The output modes for the total interaction should match the output modes of the MOPA for high-fidelity measurements.

{We can introduce quadrature operators for the frequencies of the fundamental mode
\begin{align}
x(\omega) = \tfrac{1}{\sqrt{2}} [a^\dagger(\omega) + a(\omega)], \ 
p(\omega) = \tfrac{i}{\sqrt{2}} [a^\dagger(\omega) - a(\omega)],
\end{align}
satisfying the canonical commutation relations $[x(\omega), p(\omega')] = i \delta(\omega-\omega')$. We can then describe the evolution of a set of modes in the Heisenberg picture in terms of the propagator for which we discretize the frequencies by writing $x_i = x(\omega_i) \Delta \omega, p_i =  p(\omega_i) \Delta \omega$, where $\omega_i$ are equidistant frequencies covering the region of the $(\omega,\omega')$ space where the JSA is concentrated). The operators with discretized frequencies satisfy Kronecker delta commutation relations and can be conveniently arranged in a vector
\begin{align}
	\bm{r}^T = [x_1,\ldots,x_l,p_1\ldots,p_l].
\end{align}
Their Heisenberg picture transformation is mediated by a real Symplectic matrix $\bm{S}$ which transforms the operators as follows: $\bm{r} \to \bm{r'} =\bm{S} \bm{r}.$ 
We can write symplectic matrices for both SOPA and MOPA, denoted as 
$\bm{S}_{SOPA}$, $\bm{S}_{MOPA}$, respectively. As a result, the net system is $\bm{S}_{\text{T}}=\bm{S}_{MOPA}\bm{S}_{SOPA}$. 
Using the Bloch-Messiah decomposition, we can break these propagators into terms which explicitly tell us about the input and output Schmidt modes
\begin{align}
	\bm{S}_{SOPA} &= \bm{O}_{SOPA}\bm{\Lambda}_{SOPA}\tilde{\bm{O}}^{T}_{SOPA}\\
\bm{S}_{MOPA} &= \bm{O}_{MOPA}\bm{\Lambda}_{MOPA}\tilde{\bm{O}}^{T}_{MOPA}\\
	\bm{S}_{\text{T}} &= \bm{O}_{\text{T}}\bm{\Lambda}_{\text{T}}\tilde{\bm{O}}^{T}_{\text{T}} \\
	&= \bm{O}_{MOPA}\bm{\Lambda}_{MOPA}\tilde{\bm{O}}^{T}_{MOPA}\bm{O}_{SOPA}\bm{\Lambda}_{SOPA}\tilde{\bm{O}}^{T}_{SOPA},
\end{align}
where the $\bm{O}_{\mu}, \mu\in\{SOPA,MOPA,\text{T}\}$ (which are orthogonal symplectic matrices) tell us about the output Schmidt modes, $\tilde{\bm{O}}_{\mu}$ contains information about the input Schmidt modes, and $\bm{\Lambda}_{\mu}$ (which are diagonal symplectic) describe the amount of squeezing in each stage. Since we operate the measurement and squeeze OPAs at different levels of gain, there is no guarantee that $\bm{O}_{\text{T}}=\bm{O}_{MOPA}$. 
In multimode systems, mode mixing is expected, i.e., $\tilde{\bm{O}}^{T}_{MOPA}\bm{O}_{SOPA} \neq \bm{1}$, which can produce output Schmidt modes for the full system that differ significantly from those of the measurement OPA. As a result, the observed squeezing may be reduced unless additional mode-selective techniques, such as using shaped local oscillators in balanced homodyne detection, are employed~\cite{Roslund2014, williams2025ultrafast}.}

To characterize how different the modes are, we define the fidelity between the frequency profiles of two spatiotemporal modes as
\begin{align}\label{eq:fidelity}
	F(f_{\mu},f_{\mu'}) = \left|\int d\omega \left[f_{\mu}(\omega)\right]^{*} f_{\mu'}(\omega)    \right|^{2},
\end{align}
where the labels $\mu,\mu'\in{SOPA, MOPA, T}$ and the real and imaginary parts of the mode functions $f$ are encoded in the columns of the symplectic-orthogonal matrices $O_\mu$.

In Fig.~\ref{fig:mopa}(b), we show the fidelity (Eq.~\ref{eq:fidelity}) between the output Schmidt modes of the total interaction and those of the MOPA ($F(f_{T},f_{MOPA})$) as a function of the SOPA gain (determined by the average number of signal photons generated by the SOPA $\langle  N^{SOPA}_{S}\rangle$) and for different levels of MOPA gain (in dB). The vertical red line is there as a marker to show when the SOPA generates 18 dB of gain. As demonstrated, the fidelity remains above 0.99 across the range, indicating that the output modes of the total system are, to an excellent approximation, those of the measurement OPA.

\section{Extension to the C- and L-bands}\label{sec:telecom}
{The design principles outlined in this manuscript are, in principle, applicable at any wavelength. However, not every combination of wavelength and material platform will have a corresponding geometry that meets the requirements stated previously for near-single-mode operation. In the following analysis, we allow both the ridge angle and thin-film thickness to vary. From our analysis, we find that increasing the ridge angle tends to increase the possible widths and etch parameters for which the GVM condition is met. On the other hand, increasing the film thickness tends to increase the signal GVD. }

{For FH and SH modes at 1550 nm and 775 nm respectively, TFLN platform does not naturally meet the GVD condition $\beta_{S}\gg\beta_{P}$, as the pump GVD remains higher than that of the signal at these (and shorter) wavelengths. However, starting near 1570/785 nm, we begin to observe geometries where the GVD signal overtakes the pump. Beyond this point, increasing the wavelength further increases the gap between signal and pump GVD, ultimately leading to favorable results obtained at 2090/1045 nm.}

\begin{table}[!t]
\begin{center}
\begin{tabular}{|c|c|c|c|c|}
    \hline
      Angle [$^\circ$ ] & Thickness [nm] & W [nm]& d [nm] & K  \\ \hline
      \hline
          80          &   1400        &  351      &  1232    & 1.19 \\ \hline
          90      &   1400        &  568      &  1232    & 1.19\\
    \hline
\end{tabular}
\end{center}
\caption{Optimal parameters for FH (SH) at 1570 (785) nm which meet the design principle criteria. {Due to smaller differences between the pump and signal GVD, the pump pulse duration needs to be set to 36 fs for optimal spectral purity}. Schmidt numbers are in the SPG regime.}
\label{table785}
\end{table}

{We now describe potential issues with the geometries that support near single-mode operation for the fundamental (1570 nm) and second-harmonic (785 nm) modes. Table~\ref{table785} shows two such geometries in the SPG limit. While these satisfy $\beta_{S}>\beta_{P}$, the difference is too small to yield a near-rotationally symmetric JSA (dominated by the $\beta_S$ term in the phase mismatch condition, ~\eqref{eq:pm}). {As such, to obtain the lowest Schmidt number possible we need to tune the pump pulse duration. We found a pump pulse duration of 36 fs to be optimal in terms of spectral purity}. However, the proposed geometries require steep ridge angles and narrow top widths, which are achievable with state-of-the-art fabrication processes, albeit challenging particularly for low losses.}

\begin{table}[!t]
\begin{center}
\begin{tabular}{|c|c|c|c|c|}
    \hline
  
      Angle [$^\circ$ ] & Thickness [nm]& W [nm]& d [nm] & K  \\ \hline
        \hline
          80          &   1000        &  488      &  832    & 1.16 \\ \hline
          80       &   1400        &  349      &  1232    & 1.16\\\hline
          90        &   1000        &  636      &  834    & 1.17 \\ \hline
          90        &   1400        &  566      &  1232    & 1.15\\\hline
\end{tabular}
\end{center}
\caption{Optimal parameters for FH (SH) at 1600 (800) nm, meeting the design principle criteria. {Due to smaller differences between the pump and signal GVD, the pump pulse duration needs to be set to 36 fs for optimal spectral purity}. Shown Schmidt numbers are in the SPG regime.}
\label{table800}
\end{table}

\begin{figure*}[!ht]%
	\centering
	\includegraphics[width = 1\linewidth]{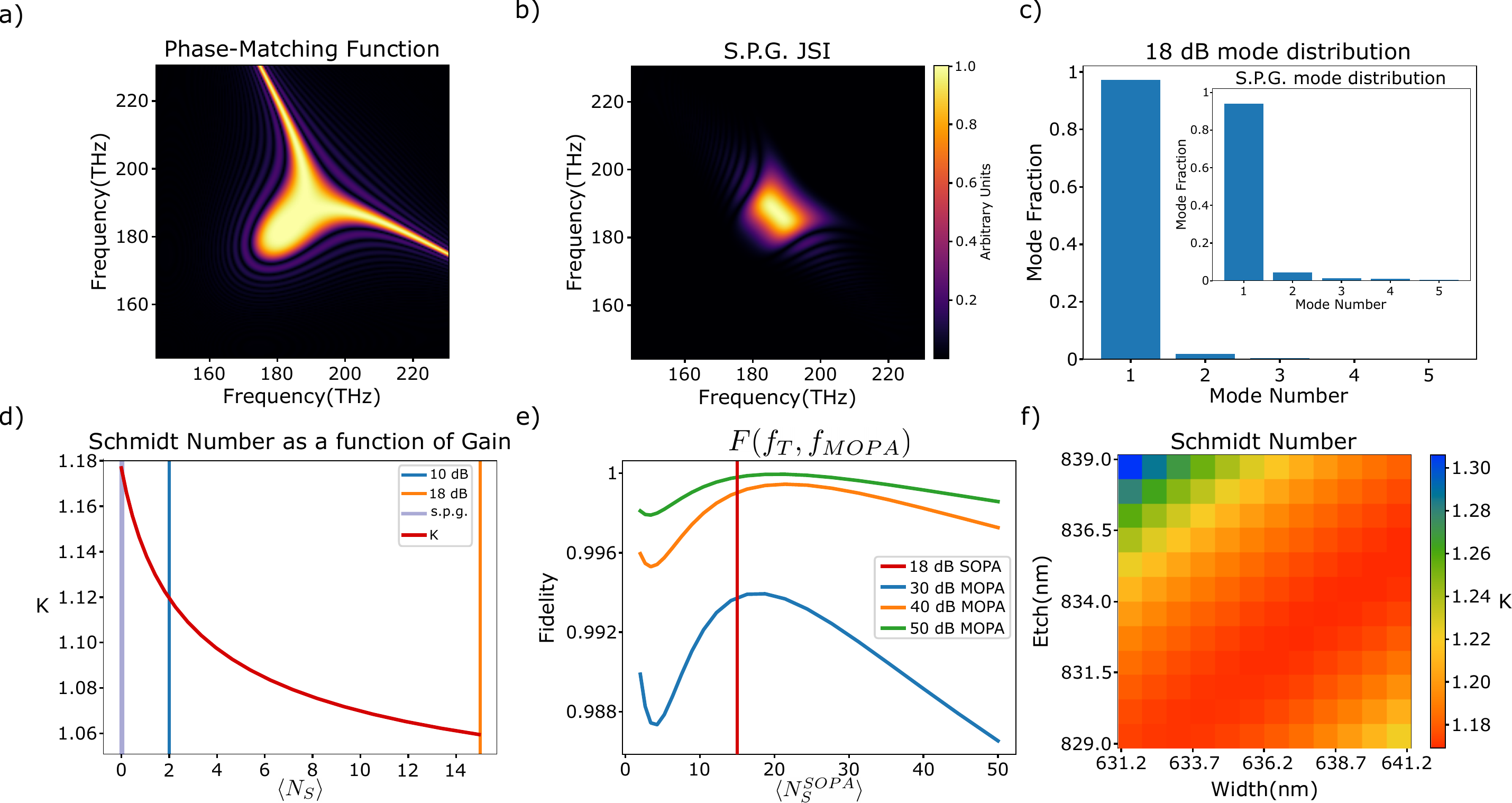}
	\caption{Design principle analysis for FH and SH modes at 1600 nm and 800 nm. Geometric parameters: film thickness=1000 nm, Etch depth = 834 nm, angle = 90$^\circ$, Width = 363.2 nm. (a) PMF for the given parameters where GVM condition is met and GVD condition is partially met. (b) JSI in the SPG regime when we optimize pump pulse duration to 36 fs.  (c) Mode distribution for 18 dB gain (inset: s.p.g. regime) with first/second mode ratio of 0.97/0.018(0.92/0.044). (d) Schmidt number as a function of average signal photon number; varies from 1.175 to 1.06. (e)Fidelity (c.f. Eq.~\ref{eq:fidelity}) between the output temporal mode of the total system ($f_{T}$) and the output temporal mode of the MOPA ($f_{MOPA}$) as a function of the level of gain in the SOPA, which is determined by the average number of signal photons generated in the SOPA $\langle N^{SOPA}_{S}  \rangle$, for different levels of gain in the MOPA (different curves). The red vertical line represents the point where the SOPA operates at 18 dB of gain. (f) Sweeping over Etch and Width near ideal values, here instead we consider variations of $\pm 5$ nm since the Schmidt number is more susceptible to variations in the geometry.}
	\label{fig:800}
\end{figure*}

{We then consider FH and SH modes at 1600 nm and 800 nm. Results are presented in Table~\ref{table800}. Although slightly offset from the telecom C-band, these wavelengths are well supported by readily available laser sources, amplifiers, modulators, and detectors that are commonly used in quantum photonic systems. Focusing on the parameters of the third row of Table~\ref{table800}, we reproduce a similar analysis as for 2090/1045 nm. In Fig.~\ref{fig:800}(a), we show
PMF for FH (1600 nm) and SH (800 nm) modes.
Because the GVD mismatch between signal and pump is modest ($\beta_S \approx 250~\text{fs}^2/\text{mm}$, $ \beta_P \approx 180~\text{fs}^2/\text{mm}$), the PMF exhibits a complex structure.
Luckily, we can optimize the pump pulse duration, setting it to 36 fs to obtain the single-photon pair JSA whose absolute value, the joint spectral intensity, is shown in Fig.~\ref{fig:800}(b). Such pulsed sources are within the reach of current laser technology~\cite{kaneshima2024development}.
With these parameters we can still achieve near-single-mode operation as shown in Fig.~\ref{fig:800}(d) where we show the Schmidt number as a function of the average signal photon number which now varies from $K\approx 1.175$ to $K\approx 1.06$. Fig.~\ref{fig:800}(c) shows the mode occupation fraction for the first five temporal modes at 18 dB of gain (inset: SPG regime). Since the Schmidt numbers are slightly higher overall, we find a first/second mode ratio of 0.97/0.018 (0.92/0.044).}

{In Fig.~\ref{fig:800}(e) we show the fidelity between the Schmidt modes of the total interaction and those of the MOPA as a a function of the SOPA gain. The behaviour is more susceptible to the differences in gain of both the SOPA and MOPA. Furthermore, the fidelity goes slightly below 0.99. This discrepancy comes from the level of dispersion present in this geometry. Indeed, in this case we find a dispersion length $L_{D}\approx 6.67$~\cite{AgrawalBook} mm which is smaller than the length of the waveguide. As such, dispersion effects need to be taken into account. For proper single-mode operation, one must therefore properly compensate for this chirping (e.g. by prechirping the pump ~\cite{suda2012effects,agrawal2019nonlinear}). The fidelity analysis was done without chirp compensation, and so the sources were more multimode than stated above, giving rise to the behavior seen in Fig.~\ref{fig:800}(e). Note that for the 2090/1045 nm geometry, we find $L_{D}\approx 58$ mm which is several times longer than the length of the waveguide.}

{Finally, in Fig.~\ref{fig:800}(f) we show how variations in the width and etch depth affect the spectral purity. We find that we are much more susceptible to variations from the ideal geometry. Unlike the 2090/1045 nm case, we only consider variations of $\pm 5$ nm. Even for these smaller variations we see bigger effects of the single-photon pair Schmidt number which now varies on the order of 0.1, at which point we are no longer single-mode.}
{Although the results are not as well behaved as the 2090/1045 nm geometry, we have shown that our design principle can be applied to other wavelengths of interest.}

Hello

\section{Conclusion and Outlook}\label{sec:end}
In summary, we propose dispersion engineering principles for travelling-wave, type-0 phase-matched OPAs to generate squeezing over a single spatiotemporal mode without auxiliary filtering. In the high-gain regime, we demonstrate how such an OPA enables all-optical, loss-tolerant measurements with near-unity fidelity. We also present concrete device designs on TFLN nanophotonics, a promising platform because of its strong optical nonlinearity and  large electro-optic coefficients. Our work will enable large-scale circuits with complex functionalities for all-optical mid-IR and telecom quantum information processing, spectroscopy, and sensing applications.

\section*{Acknowledgements}
NQ and MH acknowledge support from the Ministère de l’Économie et de l’Innovation du Québec and the Natural Sciences and Engineering Research Council of Canada. This work has been funded by the European Union’s Horizon Europe Research and Innovation Programme under agreement 101070700 project MIRAQLS. RN gratefully acknowledges support from the College of Engineering, University of Massachusetts, Amherst, and KH is thankful to Japan Science and Technology (JST) Agency (Moonshot R \& D) Grant. 

\section*{Disclosures}
The authors declare no conflicts of interest.

\section*{Data availability}

Data underlying the results presented in this paper are not publicly available at this time but may be obtained from the authors upon reasonable request.

\bibliography{scibib}

\begin{thebibliography}{10}
\newcommand{\enquote}[1]{``#1''}

\bibitem{pelucchi2022potential}
E.~Pelucchi, G.~Fagas, I.~Aharonovich, \emph{et~al.},
  {\protect\JournalTitle{Nature Reviews Physics}} \textbf{4}, 194 (2022).

\bibitem{Arrazola2021}
J.~M. Arrazola, V.~Bergholm, K.~Br\'adler, \emph{et~al.},
  {\protect\JournalTitle{Nature}} \textbf{591}, 54 (2021).

\bibitem{bourassa2021blueprint}
J.~E. Bourassa, R.~N. Alexander, M.~Vasmer, \emph{et~al.},
  {\protect\JournalTitle{Quantum}} \textbf{5}, 392 (2021).

\bibitem{aghaee2025scaling}
H.~Aghaee~Rad, T.~Ainsworth, R.~Alexander, \emph{et~al.},
  {\protect\JournalTitle{Nature}} pp. 1--8 (2025).

\bibitem{maring2024versatile}
N.~Maring, A.~Fyrillas, M.~Pont, \emph{et~al.}, {\protect\JournalTitle{Nature
  Photonics}} \textbf{18}, 603 (2024).

\bibitem{psiquantum2025manufacturable}
{PsiQuantum Team}, {\protect\JournalTitle{Nature}} pp. 1--3 (2025).

\bibitem{vernon2019scalable}
Z.~Vernon \emph{et~al.}, {\protect\JournalTitle{Phys. Rev. Applied}}
  \textbf{12}, 064024 (2019).

\bibitem{kashiwazaki2023over}
T.~Kashiwazaki, T.~Yamashima, K.~Enbutsu, \emph{et~al.},
  {\protect\JournalTitle{Applied Physics Letters}} \textbf{122} (2023).

\bibitem{Nehra2022few}
R.~Nehra, R.~Sekine, L.~Ledezma, \emph{et~al.},
  {\protect\JournalTitle{Science}} \textbf{377}, 1333 (2022).

\bibitem{quesada2018gaussianA}
N.~Quesada \emph{et~al.}, {\protect\JournalTitle{Phys. Rev. A}} \textbf{98},
  043813 (2018).

\bibitem{roman2024multimode}
V.~Roman-Rodriguez, D.~Fainsin, G.~L. Zanin, \emph{et~al.},
  {\protect\JournalTitle{Physical Review Research}} \textbf{6}, 043113 (2024).

\bibitem{uren2005pure}
A.~B. U'Ren, C.~Silberhorn, K.~Banaszek, \emph{et~al.},
  {\protect\JournalTitle{Laser Physics}} \textbf{15}, 146 (2005).

\bibitem{mosley2008heralded}
P.~J. Mosley, J.~S. Lundeen, B.~J. Smith, \emph{et~al.},
  {\protect\JournalTitle{Physical Review Letters}} \textbf{100}, 133601 (2008).

\bibitem{houde2022waveguided}
M.~Houde and N.~Quesada, {\protect\JournalTitle{AVS Quantum Science}}
  \textbf{5}, 011404 (2023).

\bibitem{houde2024perfect}
M.~Houde and N.~Quesada, {\protect\JournalTitle{AVS Quantum Science}}
  \textbf{6}, 021402 (2024).

\bibitem{zhu2021integrated}
D.~Zhu, L.~Shao, M.~Yu, \emph{et~al.}, {\protect\JournalTitle{Advances in
  Optics and Photonics}} \textbf{13}, 242 (2021).

\bibitem{christ2011probing}
A.~Christ, K.~Laiho, A.~Eckstein, \emph{et~al.}, {\protect\JournalTitle{New
  Journal of Physics}} \textbf{13}, 033027 (2011).

\bibitem{arzani2018versatile}
F.~Arzani, C.~Fabre, and N.~Treps, {\protect\JournalTitle{Phys. Rev. A}}
  \textbf{97}, 033808 (2018).

\bibitem{houde2024matrix}
M.~Houde, W.~McCutcheon, and N.~Quesada, {\protect\JournalTitle{Canadian
  Journal of Physics}} \textbf{102}, 497 (2024).

\bibitem{horoshko2024few}
D.~B. Horoshko, M.~I. Kolobov, V.~Parigi, and N.~Treps,
  {\protect\JournalTitle{Optics Letters}} \textbf{49}, 4078 (2024).

\bibitem{quesada2022beyond}
N.~Quesada, L.~Helt, M.~Menotti, \emph{et~al.}, {\protect\JournalTitle{Advances
  in Optics and Photonics}} \textbf{14}, 291 (2022).

\bibitem{helt2020degenerate}
L.~Helt and N.~Quesada, {\protect\JournalTitle{Journal of Physics: Photonics}}
  \textbf{2}, 035001 (2020).

\bibitem{Quesada2020-pdc}
N.~Quesada, G.~Triginer, M.~D. Vidrighin, and J.~E. Sipe,
  {\protect\JournalTitle{Phys. Rev. A}} \textbf{102}, 033519 (2020).

\bibitem{ledezma2022intense}
L.~Ledezma, R.~Sekine, Q.~Guo, \emph{et~al.}, {\protect\JournalTitle{Optica}}
  \textbf{9}, 303 (2022).

\bibitem{boes2023lithium}
A.~Boes, L.~Chang, C.~Langrock, \emph{et~al.}, {\protect\JournalTitle{Science}}
  \textbf{379}, eabj4396 (2023).

\bibitem{vazimali2022applications}
M.~G. Vazimali and S.~Fathpour, {\protect\JournalTitle{Advanced Photonics}}
  \textbf{4}, 034001 (2022).

\bibitem{wang2018integrated}
C.~Wang, M.~Zhang, X.~Chen, \emph{et~al.}, {\protect\JournalTitle{Nature}}
  \textbf{562}, 101 (2018).

\bibitem{Rao2019}
A.~Rao, K.~Abdelsalam, T.~Sjaardema, \emph{et~al.},
  {\protect\JournalTitle{Optics express}} \textbf{27}, 25920 (2019).

\bibitem{guo2022femtojoule}
Q.~Guo, R.~Sekine, L.~Ledezma, \emph{et~al.}, {\protect\JournalTitle{Nature
  Photonics}} \textbf{16}, 625 (2022).

\bibitem{Jankowski2020}
M.~Jankowski, C.~Langrock, B.~Desiatov, \emph{et~al.},
  {\protect\JournalTitle{Optica}} \textbf{7}, 40 (2020).

\bibitem{Juneghani2023}
F.~Arab~Juneghani, M.~Gholipour~Vazimali, J.~Zhao, \emph{et~al.},
  {\protect\JournalTitle{Advanced Photonics Research}} \textbf{4}, 2200216
  (2023).

\bibitem{snow2024}
L.~M. Ledezma, \enquote{Snow: Simulator for nonlinear optical waveguides,}
  \url{https://github.com/ledezmaluism/snow} (2024).

\bibitem{xin2025wavelength}
C.~Xin, S.~Lu, J.~Yang, \emph{et~al.}, {\protect\JournalTitle{Communications
  Physics}} \textbf{8}, 136 (2025).

\bibitem{renaud2023sub}
D.~Renaud, D.~R. Assumpcao, G.~Joe, \emph{et~al.},
  {\protect\JournalTitle{Nature Communications}} \textbf{14}, 1496 (2023).

\bibitem{sano2023effects}
Y.~Sano, Y.~Taguchi, K.~Oguchi, and Y.~Ozeki, {\protect\JournalTitle{Journal of
  the Optical Society of America B}} \textbf{41}, 183 (2023).

\bibitem{tidy3d}
{Flexcompute Inc.}, \emph{{Tidy3D}: Fast and accurate FDTD solver}, Flexcompute
  Inc. (2024). Available at \url{https://www.flexcompute.com/tidy3d/}.

\bibitem{takanashi2020all}
N.~Takanashi, A.~Inoue, T.~Kashiwazaki, \emph{et~al.},
  {\protect\JournalTitle{Optics Express}} \textbf{28}, 34916 (2020).

\bibitem{shaked2018lifting}
Y.~Shaked, Y.~Michael, R.~Z. Vered, \emph{et~al.},
  {\protect\JournalTitle{Nature communications}} \textbf{9}, 609 (2018).

\bibitem{Kalash2022}
M.~Kalash and M.~V. Chekhova, {\protect\JournalTitle{arXiv:2207.10030}}
  (2022).

\bibitem{kawasaki2024broadband}
A.~Kawasaki, R.~Ide, H.~Brunel, \emph{et~al.}, {\protect\JournalTitle{Nature
  Communications}} \textbf{15}, 9075 (2024).

\bibitem{yamashima2025all}
T.~Yamashima, T.~Kashiwazaki, T.~Suzuki, \emph{et~al.},
  {\protect\JournalTitle{Optics Express}} \textbf{33}, 5769 (2025).

\bibitem{Roslund2014}
J.~Roslund, R.~M. {de Ara{\'u}jo}, S.~Jiang, \emph{et~al.},
  {\protect\JournalTitle{Nat. Photon.}} \textbf{8}, 109 (2014).

\bibitem{williams2025ultrafast}
J.~Williams, E.~Sendonaris, R.~Nehra, \emph{et~al.},
  {\protect\JournalTitle{arXiv preprint arXiv:2502.00518}}  (2025).

\bibitem{kaneshima2024development}
K.~Kaneshima, T.~Kyoda, S.~Sugeta, and Y.~Tanaka,
  {\protect\JournalTitle{Synchrotron Radiation}} \textbf{31}, 821 (2024).

\bibitem{AgrawalBook}
G.~P. Agrawal, \enquote{Nonlinear fiber optics,} in \emph{Nonlinear Science at
  the Dawn of the 21st Century,}  (Springer, 2000), pp. 195--211.

\bibitem{suda2012effects}
A.~Suda and T.~Takeda, {\protect\JournalTitle{Applied Sciences}} \textbf{2},
  549 (2012).

\bibitem{agrawal2019nonlinear}
G.~P. Agrawal, \enquote{Optical solitons,} in \emph{Nonlinear fiber optics,}
  (Elsevier-Academic Press, 2019), 6th ed.

\end{thebibliography}

\end{document}